\documentclass[superscriptaddress,aps,twocolumn,preprintnumbers,showpacs,showkeys,nofootinbib,floatfix]{revtex4-1}
\usepackage{amssymb,amsmath,amsfonts,amsthm,graphicx,psfrag}

\newcommand{\beq}{\begin{eqnarray}}
\newcommand{\eeq}{\end{eqnarray}}

\begin{document}
\preprint{}

\title{Lattice Quantum Monte Carlo Study of Chiral Magnetic Effect in Dirac Semimetals}

\author{D.~L.~Boyda}
\email[]{boyda\_d@mail.ru}
\affiliation{Far Eastern Federal University, School of Natural Sciences, 690950 Vladivostok, Russia}

\author{V.~V.~Braguta}
\email[]{braguta@itep.ru}
\affiliation{Far Eastern Federal University, School of Biomedicine, 690950 Vladivostok, Russia}
\affiliation{Institute of Theoretical and Experimental Physics, 117259 Moscow, Russia}
\affiliation{Moscow Institute of Physics and Technology, Institutskii per. 9, Dolgoprudny, Moscow Region, 141700 Russia}
\affiliation{Institute for High Energy Physics NRC "Kurchatov Institute", Protvino, 142281 Russia}
\affiliation{Bogoliubov Laboratory of Theoretical Physics, Joint Institute for Nuclear Research, Dubna, 141980 Russia}

\author{M.~I.~Katsnelson}
\email[]{m.katsnelson@science.ru.nl}
\affiliation{Radboud University, Institute for Molecules and Materials,
Heyendaalseweg 135, NL-6525AJ Nijmegen, The Netherlands}
\affiliation{Ural Federal University, Theoretical Physics and Applied Mathematics Department, Mira Str. 19, 620002 Ekaterinburg, Russia}

\author{A.~Yu.~Kotov}
\email[]{kotov@itep.ru}
\affiliation{Institute of Theoretical and Experimental Physics, 117259 Moscow, Russia}
\affiliation{Moscow Institute of Physics and Technology, Institutskii per. 9, Dolgoprudny, Moscow Region, 141700 Russia}
\affiliation{Bogoliubov Laboratory of Theoretical Physics, Joint Institute for Nuclear Research, Dubna, 141980 Russia}

\begin{abstract}
In this paper Chiral Magnetic Effect (CME) in Dirac semimetals is studied by means of lattice Monte Carlo simulation. 
We measure conductivity of Dirac semimetals as a function of external magnetic field in parallel $\sigma_{\parallel}$ and perpendicular $\sigma_{\perp}$ to the external field directions. The simulations are carried out in three regimes: semimetal phase, onset of the insulator phase
and deep in the insulator phase. In the semimetal phase  $\sigma_{\parallel}$ grows 
whereas $\sigma_{\perp}$ drops with magnetic field. Similar behaviour was observed in the
onset of the insulator phase but conductivity is smaller and its dependence on magnetic field is weaker. Finally in the insulator 
phase conductivities $\sigma_{\parallel, \perp}$ are close to zero and do not depend on magnetic field. 
In other words, we observe manifestation of the CME current in the semimetal phase, 
weaker manifestation of the CME in the onset of the insulator phase. We do not observe signatures of CME in the 
insulator phase. We believe that the suppression of the CME current in the insulator phase is 
connected to chiral symmetry breaking and generation of dynamical fermion mass which take place in this phase.
\end{abstract}



\maketitle

Anomalies are fundamental objects in relativistic quantum field theory.
There are a lot of manifestations of the quantum anomalies in high energy physics \cite{Shifman:1988zk}. 
One of the example of the anomaly based phenomena is the Chiral Magnetic Effect(CME)
\cite{Kharzeev:2013ffa, Kharzeev:2012ph, Kharzeev:2015znc}. 
The essence of this phenomenon is generation of nondissipative electric current along external 
magnetic field in systems with the imbalance between the number of 
right-handed and left-handed fermions. One believes that the CME was observed 
in heavy ion collision experiments RHIC and LHC through the measurements of fluctuations in hadron charge asymmetry
with respect to the reaction plane\cite{Abelev:2009ac, Abelev:2012pa}.

Recent discovery of Dirac\cite{Liu864,Neupane2014,PhysRevLett.113.027603} and Weyl Semimetals\cite{Xu613,Xue1501092} 
opens the possibility to study relativistic quantum field theory phenomena 
in condensed matter physics. Characteristic feature of these materials is that the low energy fermionic spectrum
is similar to massless 3D Dirac fermions, what allows to observe different manifestations of the quantum anomalies and,
in particular, the CME. 

To observe the CME, it is necessary to create system with imbalance between the number of right-handed and left-handed 
fermions. This can be done if one applies parallel electric $\bf E$ and magnetic $\bf B$ fields to the system,
what leads to generation of the density of chiral charge with the rate\cite{Li:2014bha}
\beq
\frac {d \rho_5} {d t} = \frac {e^2} {4 \pi^2 \hbar^2 c} {\bf E}\cdot {\bf B} - \frac {\rho_5} {\tau}.
\eeq 
The first term in last equation describes the production of chiral charge due to the chiral anomaly, while the second one describes 
the decrease of chirality due to the chirality-changing processes. Note that we use Lorentz-Heaviside units throughout the paper. The $\tau$ is the relaxation time of chiral charge which 
was studied in \cite{Manuel:2015zpa, Ruggieri:2016asg}.
At large times as the result of the balance between production due to the anomaly and 
decrease due to the chirality-changing processes, the system stabilizes at the chiral charge density given by the formula
\beq
\rho_5 = \frac {e^2} {4 \pi^2 \hbar^2 c} {\bf E}\cdot {\bf B} \tau
\label{rho5}
\eeq  
The chiral charge density can be parameterized by the chiral chemical potential $\mu_5$ through 
the equation of state (EoS) $\rho_5=\rho_5(\mu_5)$. Below we are going to use the linear response 
theory for which the electric field $\bf E$ is considered as a perturbation. In this limit 
one can state that the chiral chemical potential created in the system is small. 
For the small chiral chemical potential the EoS can be written as
\beq
\rho_5=\mu_5 \chi(T, B)
+ O(\mu_5^3),
\label{eof}
\eeq  
where the $\chi(T, B)$ is a function of magnetic field and temperature. 
It is clear that in the limit of small magnetic field $eB/T^2 \to 0$  the behaviour 
of the function $\chi$ is determined by temperature and the $\chi(T, B) \sim T^2$.
In the limit of large magnetic field $eB/T^2 \to \infty$ the function $\chi$ is determined 
by degeneracy on the lowest Landau level and one can expect that $\chi(T, B) \sim eB$. As was noted above the influence of the external magnetic field 
on the system with chiral imbalance leads to generation of electric current which is given by 
the formula
\beq
{\bf j_{CME}}= \frac {e^2} {2 \pi^2} \mu_5 {\bf B}.
\label{cme}
\eeq
Combining formulae (\ref{rho5}), (\ref{eof}) and ({\ref{cme}}) one acquires conductivity due to the CME
\beq
j_{CME}^i = \sigma_{CME}^{ij} E^{j},~ \sigma_{CME}^{ij}= \frac {e^4} {8 \pi^4 \hbar^2 c} \frac {\tau} {\chi(T,B)} B^{i} B^{k} 
\label{cme_cond}
\eeq 
Below it will be assumed that the magnetic field is directed along $z$ axis. 

In addition to the CME current there is also Ohmic current. Total conductivity 
is the sum of Ohmic and the CME conductivities $\sigma=\sigma^{O}+\sigma^{CME}$.
If electric field is applied along $x$ axis(perpendicular to the magnetic field),
the Lorentz force acts to charged particles leading to decrease of the $\sigma^{O}_{xx}$ 
component, i.e. positive magnetoresistance. 
The $\sigma^{CME}_{xx}$ component is zero in this case. 
On the other hand if electric field is applied along magnetic field, 
there is no Lorentz force and magnetoresistance of the $\sigma^{O}_{zz}$ component. 
At the same time the $\sigma^{CME}_{zz}$ is rising function of the magnetic field. 
So, the growth of the $\sigma_{zz}$ with magnetic field or negative magnetoresistance is a signature of the CME. 
From formula (\ref{cme_cond}) and the properties of the function $\chi(T,B)$ one can see that for the small magnetic field 
the $\sigma^{CME}_{zz}$ rises quadratically with the magnetic field.
For the sufficiently large magnetic field quadratically rising function 
switches to linearly rising behaviour. Experimental observation of the CME in Dirac 
semimetals through the measurement of the conductivity was reported in \cite{Li:2014bha,Li2015,Li2016}. 
Notice also that the CME in QCD in the quenched approximation was studied in paper \cite{Buividovich:2010tn}.

It is known that due to the smallness of the Fermi velocity the interactions between 
quasiparticles in Dirac semimetals are strong what can 
lead to a considerable modification of the above formulae. Moreover, it is 
known that large coupling constant can lead to the dynamical chiral symmetry breaking and 
the transition from semimetal to insulator phase \cite{Braguta:2016vhm, Braguta:2017voo}.
For the CME the chiral symmetry of the fermionic sector of the theory is important. 
In this paper we are going to address the question how the interactions between fermions 
influence the CME in Dirac semimetals. In particular, we are going to study how the transition to the phase with the
dynamical chiral symmetry breaking modifies the CME. In this paper we are going to apply 
quantum Monte Carlo simulations\cite{Montvay:1994cy} which fully take into account many-body effects in Dirac semimetals for an arbitrary coupling constant $\alpha_{eff}$. This approach is successfully used to study various strongly correlated condense matter systems \cite{Drut:2008rg, Drut:2009aj,Hands:2008id,Hands:2008id,Ulybyshev:2013swa,DeTar:2016dmj,DeTar:2016vhr, Buividovich:2012nx,Boyda:2016emg, Yamamoto:2016rfr,Yamamoto:2016zpx, Braguta:2016vhm,Braguta:2017voo}.

To simplify our study we are going to carry out Monte Carlo simulation of Dirac semimetals
with two Fermi points and small isotropic Fermi velocity $v_F \ll c$. Low energy effective theory of 
fermionic excitations can be described by two flavors of 3D Dirac fermions. 
Due to the smallness of the Fermi velocity magnetic interactions and retardation
effects can be safely disregarded. As the result the interaction in Dirac semimetals is reduced to instantaneous
Coulomb potential with the effective coupling constant $\alpha_{eff}=\alpha_{el} \cdot c / v_F$, 
where $\alpha_{el}=e^2/(4\pi\hbar c)=1/137$.

The partition function of the system under study can be written as
\begin{equation}
	Z=\int D\psi D\bar{\psi}DA_4 \exp(-S_E),
\label{eq:partfunc1}
\end{equation}
where $\bar{\psi}$, $\psi$ are fermion fields and the action $S_E$ is
\begin{equation}
\begin{split}
	S_E=\sum\limits_{a=1}^2\int d^3xdt\bar{\psi}_a(\gamma_4(\partial_4+iA_4)+\gamma_i(\partial_i+iA_i))\psi_a+\\
	+\frac{1}{8\pi\alpha_{eff}}\int d^3xdt(\partial_iA_4)^2
\label{eq:partfunc}
\end{split}
\end{equation}
Note that in formulae (\ref{eq:partfunc1}),~(\ref{eq:partfunc}) the interactions between quasiparticles are transmitted by the field $A_4$ whereas 
the vector part of the gauge potential $A_i$ is introduced in order to describe the external magnetic field.

To write a discretized version of action (\ref{eq:partfunc}) 
we introduce a regular cubic lattice in four dimensional space with spatial lattice spacing 
$a_s$ and temporal lattice spacing $a_t$ ($a_t=\xi a_s$). As discussed in \cite{Braguta:2017voo}, it is important to take the limit $\xi=a_t/a_s\to0$. The number of lattice sites is $L_s$ in each spatial direction 
and $L_t$ in temporal direction. 
Below we will take $a_s=1$, restoring explicit spatial lattice 
spacing when necessary.

In our simulations we use staggered fermions coupled to Abelian lattice gauge field $\theta_4(x)$.
The Euclidean discretized actions for the fermion fields $S_f$ and gauge fields $S_g$ can be written as
\begin{equation}
\begin{split}
S_f=\bar{\Psi}_xD_{x,y}\Psi_y=\sum\limits_x\left(m\bar{\Psi}_x\Psi_x+\right.\\
\left.
+\frac12[\bar{\Psi}_x\eta_4(x)e^{i\theta_4(x)}\Psi_{x+\hat{4}}-\bar{\Psi}_{x+4}\eta_4(x)e^{-i\theta_4(x)}\Psi_x]\right.+\\
+\frac12\sum\limits_{i=1}^3\xi [\bar{\Psi}_x\eta_i(x)e^{i\theta_i(x)}\Psi_{x+\hat{\i}}-\bar{\Psi}_{x+\i}\eta_i(x)e^{-i\theta_i(x)}\Psi_x]
\left.\right).\\
S_g=\frac{\beta}{2 \xi}\sum_{x,i}(\theta_4(x)-\theta_4(x+i))^2,\quad \beta=\frac 1 { 4 \pi \alpha_{eff}} \label{eq:latticeaction}
\end{split}
\end{equation}
where $\eta_{\mu}(x)=(-1)^{x_1+\ldots+x_{\mu-1}},\mu=1,\ldots,4$.
The lattice field $\theta_4$ is related to the continuum Abelian field $A_4$ as $\theta_4=a_t A_4$. 
It should be noted that nonzero mass term in (\ref{eq:latticeaction}) is
necessary in order to ensure the invertibility of the staggered Dirac operator $D_{x,y}$. 

\begin{figure}[t]
\includegraphics[scale=0.6,clip=false]{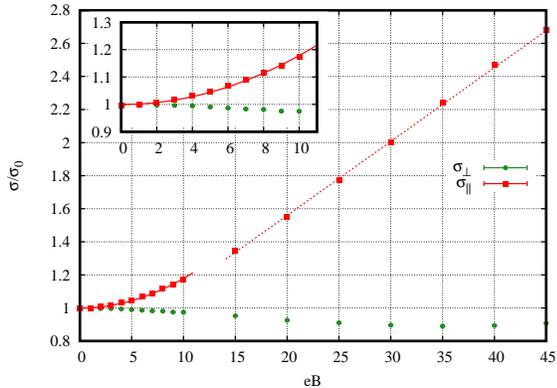}
\caption{The ratios $\sigma_{\parallel, \perp}/\sigma_0$ as a function of the external magnetic field $eB$. 
The system is in the semimetal phase with the $\alpha_{eff}=0.32$. 
The $\sigma_0$ is the conductivity at $\alpha_{eff}=0.32$ at zero magnetic field.
The external magnetic field $eB$ is shown in units $\frac{2\pi}{(a_s L_s)^2}$.
Red points correspond to conductivity parallel to the magnetic field $\sigma_{\parallel}$, green points give conductivity in the perpendicular direction $\sigma_{\perp}$ . Red lines correspond to quadratic and linear fit.}
\label{fig:conductivity1}
\end{figure} 

In order to get the theory for two Dirac fermions we numerically 
take square root of the fermion determinant. A detailed description of lattice action 
(\ref{eq:latticeaction}) and rooting procedure can be found in paper \cite{Braguta:2017voo}. 

The fields $\theta_i$ describe the external magnetic field. Explicit expressions for the fields
$\theta_i$ can be found in \cite{AlHashimi:2008hr}. It is important to notice here that 
due to the periodic boundary conditions on the $\theta_i$ magnetic field on the lattice
is quantized
\begin{equation}
	eB=\frac{2\pi N_b}{(a_s L_s)^2}, \qquad N_b\in \mathbb{Z}
\end{equation}
The temperature on the lattice is given by the expression $T=1/(L_t a_t)$. 
So, for the isotropic lattice $L_s=L_t$,~$a_s=a_t$  the ratio $eB/T^2=2 \pi N_b>1$,
what does not allow to study the small magnetic field. In order to study 
the influence of the small magnetic field on the system we use anisotropic lattice with 
different lattice steps in temporal and spatial directions $a_t=\xi a_s$ with $\xi=1/4$.

To determine the electric conductivity we measured the Euclidean correlator $C_i(t)$ of the spatial components of electric current $j_{i}=\bar{\psi}\gamma_{i}\psi,~ i=1,2,3$ (no summation over $i$ is assumed):
\begin{equation}
	C_i(t)=\int d^3\bar{x} \langle j_{i}({\bar x},t)j_{i}(0)\rangle
\label{eq:corr}
\end{equation}
In lattice simulations we used conserved current for staggered fermions\cite{DeGrand:2006zz}:
\begin{equation}
	j_i(x)=\frac14(\bar{\Psi}_{x+i}\eta_i(x)U^{\dag}_i(x)\Psi_x+\bar{\Psi}_{x}\eta_i(x)U_i(x)\Psi_{x+i})
\end{equation}
Notice that in the continuum limit $a_s \to 0$ this current corresponds to usual electric current 
of 3D Dirac fermions $\bar{\psi}\gamma_{i}\psi$.

\begin{figure}[t]
\includegraphics[scale=0.6,clip=false]{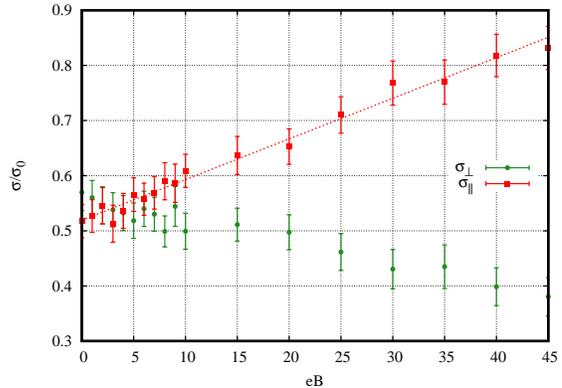}
\caption{Same as in Fig.~\ref{fig:conductivity1}, but for $\alpha_{eff}=1.27$ where 
the system is in the onset of the insulator phase.
}
\label{fig:conductivity2}
\end{figure}

The correlator (\ref{eq:corr}) is related to the conductivity $\sigma_{ii}(\omega)$ by means of Kubo relation\cite{Buividovich:2012nx}:
\begin{equation}
	C_i(t)=\int\limits_0^{\infty}\frac{d\omega}{2\pi}K(\omega,t)\sigma_{ii}(\omega),
\label{kubo}
\end{equation}
where $K(\omega,t)=\frac{2\omega \cosh(w(\frac{1}{2T}-t))}{\sinh(\frac{\omega}{2T})}$ is the standard thermal kernel.
It should be noticed that the formula for the conductivity (\ref{cme_cond}) was derived from real time equation. 
At the same time in lattice quantum Monte Carlo simulations we study the system in the thermodynamical equilibrium where 
there is no real time. In real time dynamics the electrical conductivity is related to the retarded Green function of the 
electromagnetic currents through Kubo relation. In turn the retarded Green function is related to the 
Euclidean correlation function of the electromagnetic currents (\ref{eq:corr}). For this reason one can use the 
Euclidean correlation fuctions in order to study real time transport coefficients (see formula (\ref{kubo})). This approach is used in our paper.

In order to study the CME one needs the conductivity $\sigma_{ii}(\omega)$ at small frequencies. 
To calculate it one has to invert equation (\ref{kubo}) what turns out to be very complicated problem. Typically there are several tens of points for $C_i(t)$ whereas one has to determine the continuous function 
$\sigma_{ii}(\omega)$. 

\begin{figure}[t]
\includegraphics[scale=0.6,clip=false]{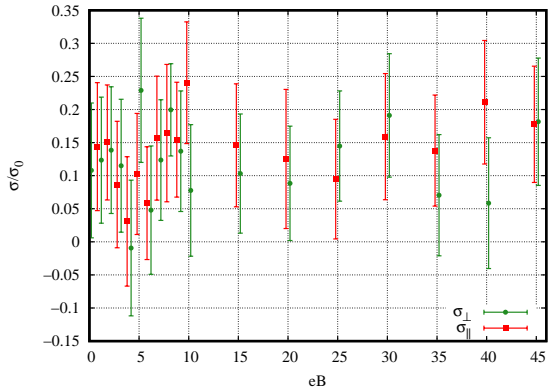}
\caption{Same as in Fig.~\ref{fig:conductivity1}, but for $\alpha_{eff}=2.55$
where the system is deep in the insulator phase. In order to avoid superposition of data points for $\sigma_{\parallel}$ and $\sigma_{\perp}$ we applied a tiny shift along the $eB$ axis.
}
\label{fig:conductivity3}
\end{figure}

In this paper we are going to use midpoint calculation of the conductivity\cite{Buividovich:2012nx}:
\begin{equation}
	\bar {\sigma}_{ii}=\int\limits_0^{\infty}\frac{d\omega}{2\pi}\frac{2\omega}{\sinh\frac{\omega}{2T}}\sigma_{ii}(\omega) / \int\limits_0^{\infty}\frac{d\omega}{2\pi}\frac{2\omega}{\sinh\frac{\omega}{2T}} = \frac{1}{\pi T^2} C_i \biggl (\frac{1}{2T} \biggr )
\label{cond}
\end{equation}
Physically the estimator $\bar {\sigma}_{ii}$ averages the conductivity over the interval $\sim \mbox{few}\times T$.

In the calculation we use the lattice with the size $L_s=L_t=20$, 
the asymmetry $\xi=a_t/a_s=1/4$ (one can expect that this value is close to the limit $\xi\to0$\cite{Braguta:2017voo}) and the fermion mass in lattice units $m=0.0025$. Lattice simulation are carried out for the effective coupling constant $\alpha_{eff}=2.55, 1.27, 0.32$. At $\alpha_{eff}^c=1.14$ there is transition from the semimetal to the insulator phase with dynamical breaking of the chiral symmetry\cite{Braguta:2017voo}. 
So the system with $\alpha_{eff}=0.32$ is in the semimetal phase. The $\alpha_{eff}=1.27$ 
corresponds to the insulator phase but close to the transition point $\alpha_{eff}^c$
(the onset of the insulator phase).  
The system with $\alpha_{eff}=2.55$ is deep in the insulator phase. For each value of the $\alpha_{eff}$
the simulations are conducted for a set of values of external magnetic field. For this set of parameters
we calculated midpoint conductivity perpendicular $\sigma_{\perp} = \bar {\sigma}_{xx}$
and parallel $\sigma_{\parallel} = \bar {\sigma}_{zz}$ to the applied magnetic field. 

In Fig.~\ref{fig:conductivity1},~\ref{fig:conductivity2},~\ref{fig:conductivity3} we plot the ratios $\sigma_{\parallel, \perp}/\sigma_0$ as a function of the magnetic field $eB$ for different values of the $\alpha_{eff}$: 
$\alpha_{eff}=0.32$ (the semimetal phase, Fig.~\ref{fig:conductivity1}), $\alpha_{eff}=1.27$ (the onset of the insulator phase, Fig.~\ref{fig:conductivity2}) and $\alpha_{eff}=2.55$ (deep in the insulator phase, Fig.~\ref{fig:conductivity3}). The $\sigma_0$ is conductivity $\bar \sigma_{xx}$ at $\alpha_{eff}=0.32$ at zero magnetic field.
Red points on these figures correspond to conductivity parallel to the magnetic field $\sigma_{\parallel}$, 
green points give conductivity in the perpendicular direction $\sigma_{\perp}$.

First let us consider Fig.~\ref{fig:conductivity1} where the system is in the semimetal phase with $\alpha_{eff}=0.32$. 
It is clearly seen $\sigma_{\parallel}$ grows whereas $\sigma_{\perp}$ drops with the magnetic field.  
So we see positive magnetoresistance for the $\sigma_{\perp}$ and negative magnetoresistance
for the $\sigma_{\parallel}$ what agrees with our expectation and the experiment \cite{Li:2014bha}.
Notice that there are two regimes of the dependence of the conductivity $\sigma_{\parallel}(B)$ on the magnetic field.
For small values of magnetic field the $\sigma_{\parallel}(B)$ is quadratically rising function, 
while for larger values of magnetic field it is linearly rising function. This behaviour 
of the $\sigma_{\parallel}$ is in agreement with our expectation from formula (\ref{cme_cond}). 
This brings us to the conclusion that we observe the CME in the lattice simulation of Dirac semimetals
in the semimetal phase.   

Further let us consider Fig.~\ref{fig:conductivity2} with the $\alpha_{eff}=1.27$ where the system is in the onset of the insulator phase. 
From Fig.~\ref{fig:conductivity2} it is seen that the $\sigma_{\parallel}(B)$ and $\sigma_{\perp}(B)$
behave similarly to the semimetal phase, but their absolute values are smaller
and the CME is weaker. 

Finally in Fig.~\ref{fig:conductivity3} the results for the system deep in the insulator phase 
with $\alpha_{eff}=2.55$ are shown. It is seen that the conductivity is 
close to zero, it does not depend on the magnetic field and we do not observe the CME.    

\begin{figure}[t]
\includegraphics[scale=0.4,clip=false]{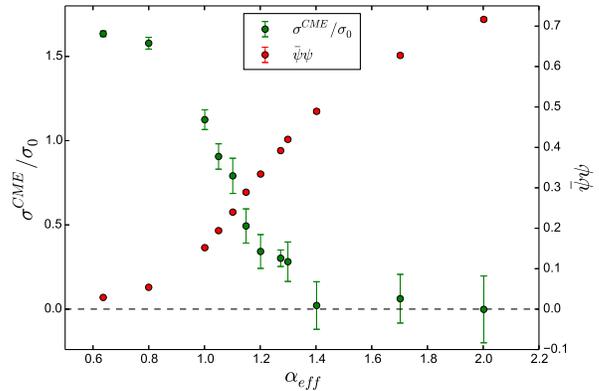}
\caption{The left $y$-axis is the ratio of the CME conductivity $\sigma^{CME}$ for the magnetic 
field $eB (a_s L_s)^2/(2\pi)=40$ to the $\sigma_0$ (green points)
as a function of the effective coupling constant $\alpha_{eff}$.
The right $y$-axis is the chiral condensate $\bar \psi \psi$ (red points)  for the $eB (a_s L_s)^2/(2\pi)=40$
as a function of the $\alpha_{eff}$.
}
\label{fig:conductivity4}
\end{figure}

In order to get more insight about how the CME is affected by the chiral symmetry breaking in Dirac semimetals 
we measured the CME conductivity $\sigma^{CME}$ and the chiral condensate $\bar \psi \psi$ 
for the magnetic field $eB (a_s L_s)^2/(2\pi)=40$ for different values of the effective coupling constant $\alpha_{eff}$. 
In Fig.~\ref{fig:conductivity4} we plot the ratio of the CME conductivity $\sigma^{CME}$ to the 
$\sigma_0$ (green points) as a function of the effective coupling constant $\alpha_{eff}$.
On the same figure we also plot the chiral condensate $\bar \psi \psi$ (red points) as a function of the $\alpha_{eff}$. 

To understand the physical meaning of Fig.~\ref{fig:conductivity4} let us recall that the chiral 
condensate is sensitive to the dynamical chiral symmetry breaking/restoration transition. In the region 
$\alpha_{eff}<1$ the chiral condenstate is small and the chiral symmetry is not dynamically 
broken. In the region $\alpha_{eff}>1.4$ the chiral consensate is large and the chiral symmetry is
dynamically broken. In the region $1<\alpha_{eff}<1.4$ we see rapid rise of the chiral condensate
what implies that this is transition region between these two regimes. The critical coupling 
constant for this transition is $\alpha_{eff}^c\simeq1.1$. Now let us consider the dependense of the 
CME conductivity $\sigma^{CME}$ on the $\alpha_{eff}$. It is seen from Fig.~\ref{fig:conductivity4} 
that in the region $\alpha_{eff}<1$ the $\sigma^{CME}$ weakly depends on the $\alpha_{eff}$. 
However, for the $\alpha_{eff}>1$ we observe rapid decrease of the conductivity. 
Finally in the region $\alpha_{eff}>1.4$ the CME conductivity is zero within the uncertainty of the 
calculations. Notice that the region where the chiral condensate rises coincides with the region 
where the CME conductivity drops. This confirms that the CME conductivity is very 
sensitive to the chiral symmetry and dynamical breaking of the chiral symmetry 
in the system.

To summarize, the CME can be well seen in the semimetal phase where the chiral symmetry is not dynamically broken. 
It is also seen that the CME conductivity rapidly drops in the region where there is the transition from 
the phase with chiral symmetry to the phase where the chiral symmetry is dynamically broken. 
Finally in the region where the system is deep in the insulator phase with 
dynamically broken chiral symmetry the CME conductivity is zero within the uncertainty of the 
calculation. This behaviour of the conductivity can be understood as follows. 
Generation of the CME current is connected with the Schwinger pair production on the lowest Landau level\cite{Kharzeev:2013jha}. 
In the insulator phase due to chiral symmetry breaking there is dynamical generation 
of the fermion mass whereas in the semimetal phase there is no dynamical fermion mass. 
The deeper to the insulator phase the larger the dynamical fermion mass. 
Note also that the larger the fermion mass the larger suppression of the Schwinger pair production. 
For this reason one can expect that the CME is suppressed in the onset of the 
insulator phase as compared to the semimetal phase. It is also reasonable 
to assume that there is no CME deep to the insulator phase since 
the dynamical fermion mass is large and the Schwinger pair production at the lowest Landau level
is considerably suppressed.

In conclusion, in this paper the Chiral Magnetic Effect in Dirac
semimetals was studied by means of
lattice Monte Carlo simulation.  The simulations were carried out in
three regimes: semimetal phase,
onset of the insulator phase and deep in the insulator phase.
We observe manifestation of the CME current in the semimetal phase,
weaker manifestation of the CME in the onset of the insulator phase.
We do not observe the CME in the
insulator phase.

In order to get more insight about how the CME is affected by the
chiral symmetry breaking we measured the
CME conductivity and the chiral condensate at large magnetic field as
a function of the effective coupling
constant $\alpha_{eff}$. We found that the large the chiral condensate
the smaller the CME conductivity.
Thus we confirmed that the CME conductivity is very sensitive to the
chiral symmetry and dynamical
breaking of the chiral symmetry in the system.

\begin{acknowledgments}
We would like to thank D.~Kharzeev, M.~Chernodub and N.~Astrakhantsev for useful discussions.
The work of MIK was supported by Act 211 Government of the Russian
Federation, Contract No. 02.A03.21.0006. The work of VVB and AYK, which consisted of
numerical simulation and calculation of the conductivity,
was supported by grant from the Russian Science Foundation (project number 16-12-10059).
The work of DLB was supported by RFBR Grant No. 16-32-00362-mol-a.
This work has been partly carried out using computing resources of the
federal collective usage center Complex for Simulation and Data
Processing for Mega-science Facilities at NRC “Kurchatov Institute”,
http://ckp.nrcki.ru/. The authors are also grateful for the provided computer resources by Far East Computing Resource "Far Eastern Computing Resource" equipment (https://www.cc.dvo.ru).
\end{acknowledgments}

\bibliographystyle{apsrev4-1}

\begin{thebibliography}{32}%
\makeatletter
\providecommand \@ifxundefined [1]{%
 \@ifx{#1\undefined}
}%
\providecommand \@ifnum [1]{%
 \ifnum #1\expandafter \@firstoftwo
 \else \expandafter \@secondoftwo
 \fi
}%
\providecommand \@ifx [1]{%
 \ifx #1\expandafter \@firstoftwo
 \else \expandafter \@secondoftwo
 \fi
}%
\providecommand \natexlab [1]{#1}%
\providecommand \enquote  [1]{``#1''}%
\providecommand \bibnamefont  [1]{#1}%
\providecommand \bibfnamefont [1]{#1}%
\providecommand \citenamefont [1]{#1}%
\providecommand \href@noop [0]{\@secondoftwo}%
\providecommand \href [0]{\begingroup \@sanitize@url \@href}%
\providecommand \@href[1]{\@@startlink{#1}\@@href}%
\providecommand \@@href[1]{\endgroup#1\@@endlink}%
\providecommand \@sanitize@url [0]{\catcode `\\12\catcode `\$12\catcode
  `\&12\catcode `\#12\catcode `\^12\catcode `\_12\catcode `\%12\relax}%
\providecommand \@@startlink[1]{}%
\providecommand \@@endlink[0]{}%
\providecommand \url  [0]{\begingroup\@sanitize@url \@url }%
\providecommand \@url [1]{\endgroup\@href {#1}{\urlprefix }}%
\providecommand \urlprefix  [0]{URL }%
\providecommand \Eprint [0]{\href }%
\providecommand \doibase [0]{http://dx.doi.org/}%
\providecommand \selectlanguage [0]{\@gobble}%
\providecommand \bibinfo  [0]{\@secondoftwo}%
\providecommand \bibfield  [0]{\@secondoftwo}%
\providecommand \translation [1]{[#1]}%
\providecommand \BibitemOpen [0]{}%
\providecommand \bibitemStop [0]{}%
\providecommand \bibitemNoStop [0]{.\EOS\space}%
\providecommand \EOS [0]{\spacefactor3000\relax}%
\providecommand \BibitemShut  [1]{\csname bibitem#1\endcsname}%
\let\auto@bib@innerbib\@empty
\bibitem [{\citenamefont {Shifman}(1991)}]{Shifman:1988zk}%
  \BibitemOpen
  \bibfield  {author} {\bibinfo {author} {\bibfnamefont {M.~A.}\ \bibnamefont
  {Shifman}},\ }\href {\doibase 10.1016/0370-1573(91)90020-M} {\bibfield
  {journal} {\bibinfo  {journal} {Phys. Rept.}\ }\textbf {\bibinfo {volume}
  {209}},\ \bibinfo {pages} {341} (\bibinfo {year} {1991})},\ \bibinfo {note}
  {[Usp. Fiz. Nauk157,561(1989)]}\BibitemShut {NoStop}%
\bibitem [{\citenamefont {Kharzeev}(2014)}]{Kharzeev:2013ffa}%
  \BibitemOpen
  \bibfield  {author} {\bibinfo {author} {\bibfnamefont {D.~E.}\ \bibnamefont
  {Kharzeev}},\ }\href {\doibase 10.1016/j.ppnp.2014.01.002} {\bibfield
  {journal} {\bibinfo  {journal} {Prog. Part. Nucl. Phys.}\ }\textbf {\bibinfo
  {volume} {75}},\ \bibinfo {pages} {133} (\bibinfo {year} {2014})},\ \Eprint
  {http://arxiv.org/abs/1312.3348} {arXiv:1312.3348 [hep-ph]} \BibitemShut
  {NoStop}%
\bibitem [{\citenamefont {Kharzeev}\ \emph
  {et~al.}(2013{\natexlab{a}})\citenamefont {Kharzeev} \emph
  {et~al.}}]{Kharzeev:2012ph}%
  \BibitemOpen
  \bibfield  {author} {\bibinfo {author} {\bibfnamefont {D.~E.}\ \bibnamefont
  {Kharzeev}} \emph {et~al.},\ }\href {\doibase 10.1007/978-3-642-37305-3_1}
  {\bibfield  {journal} {\bibinfo  {journal} {Lect. Notes Phys.}\ }\textbf
  {\bibinfo {volume} {871}},\ \bibinfo {pages} {1} (\bibinfo {year}
  {2013}{\natexlab{a}})},\ \Eprint {http://arxiv.org/abs/1211.6245}
  {arXiv:1211.6245 [hep-ph]} \BibitemShut {NoStop}%
\bibitem [{\citenamefont {Kharzeev}\ \emph {et~al.}(2016)\citenamefont
  {Kharzeev} \emph {et~al.}}]{Kharzeev:2015znc}%
  \BibitemOpen
  \bibfield  {author} {\bibinfo {author} {\bibfnamefont {D.~E.}\ \bibnamefont
  {Kharzeev}} \emph {et~al.},\ }\href {\doibase 10.1016/j.ppnp.2016.01.001}
  {\bibfield  {journal} {\bibinfo  {journal} {Prog. Part. Nucl. Phys.}\
  }\textbf {\bibinfo {volume} {88}},\ \bibinfo {pages} {1} (\bibinfo {year}
  {2016})},\ \Eprint {http://arxiv.org/abs/1511.04050} {arXiv:1511.04050
  [hep-ph]} \BibitemShut {NoStop}%
\bibitem [{\citenamefont {Abelev}\ \emph {et~al.}(2009)\citenamefont {Abelev}
  \emph {et~al.}}]{Abelev:2009ac}%
  \BibitemOpen
  \bibfield  {author} {\bibinfo {author} {\bibfnamefont {B.~I.}\ \bibnamefont
  {Abelev}} \emph {et~al.} (\bibinfo {collaboration} {STAR}),\ }\href {\doibase
  10.1103/PhysRevLett.103.251601} {\bibfield  {journal} {\bibinfo  {journal}
  {Phys. Rev. Lett.}\ }\textbf {\bibinfo {volume} {103}},\ \bibinfo {pages}
  {251601} (\bibinfo {year} {2009})},\ \Eprint {http://arxiv.org/abs/0909.1739}
  {arXiv:0909.1739 [nucl-ex]} \BibitemShut {NoStop}%
\bibitem [{\citenamefont {Abelev}\ \emph {et~al.}(2013)\citenamefont {Abelev}
  \emph {et~al.}}]{Abelev:2012pa}%
  \BibitemOpen
  \bibfield  {author} {\bibinfo {author} {\bibfnamefont {B.}~\bibnamefont
  {Abelev}} \emph {et~al.} (\bibinfo {collaboration} {ALICE}),\ }\href
  {\doibase 10.1103/PhysRevLett.110.012301} {\bibfield  {journal} {\bibinfo
  {journal} {Phys. Rev. Lett.}\ }\textbf {\bibinfo {volume} {110}},\ \bibinfo
  {pages} {012301} (\bibinfo {year} {2013})},\ \Eprint
  {http://arxiv.org/abs/1207.0900} {arXiv:1207.0900 [nucl-ex]} \BibitemShut
  {NoStop}%
\bibitem [{\citenamefont {Liu}\ \emph {et~al.}(2014)\citenamefont {Liu} \emph
  {et~al.}}]{Liu864}%
  \BibitemOpen
  \bibfield  {author} {\bibinfo {author} {\bibfnamefont {Z.~K.}\ \bibnamefont
  {Liu}} \emph {et~al.},\ }\href {\doibase 10.1126/science.1245085} {\bibfield
  {journal} {\bibinfo  {journal} {Science}\ }\textbf {\bibinfo {volume}
  {343}},\ \bibinfo {pages} {864} (\bibinfo {year} {2014})}\BibitemShut
  {NoStop}%
\bibitem [{\citenamefont {Neupane}\ \emph {et~al.}(2014)\citenamefont {Neupane}
  \emph {et~al.}}]{Neupane2014}%
  \BibitemOpen
  \bibfield  {author} {\bibinfo {author} {\bibfnamefont {M.}~\bibnamefont
  {Neupane}} \emph {et~al.},\ }\href {http://dx.doi.org/10.1038/ncomms4786}
  {\bibfield  {journal} {\bibinfo  {journal} {Nat. Commun.}\ }\textbf {\bibinfo
  {volume} {5}},\ \bibinfo {pages} {3786 EP } (\bibinfo {year}
  {2014})}\BibitemShut {NoStop}%
\bibitem [{\citenamefont {Borisenko}\ \emph {et~al.}(2014)\citenamefont
  {Borisenko} \emph {et~al.}}]{PhysRevLett.113.027603}%
  \BibitemOpen
  \bibfield  {author} {\bibinfo {author} {\bibfnamefont {S.}~\bibnamefont
  {Borisenko}} \emph {et~al.},\ }\href {\doibase
  10.1103/PhysRevLett.113.027603} {\bibfield  {journal} {\bibinfo  {journal}
  {Phys. Rev. Lett.}\ }\textbf {\bibinfo {volume} {113}},\ \bibinfo {pages}
  {027603} (\bibinfo {year} {2014})}\BibitemShut {NoStop}%
\bibitem [{\citenamefont {Xu}\ \emph {et~al.}(2015{\natexlab{a}})\citenamefont
  {Xu} \emph {et~al.}}]{Xu613}%
  \BibitemOpen
  \bibfield  {author} {\bibinfo {author} {\bibfnamefont {S.-Y.}\ \bibnamefont
  {Xu}} \emph {et~al.},\ }\href {\doibase 10.1126/science.aaa9297} {\bibfield
  {journal} {\bibinfo  {journal} {Science}\ }\textbf {\bibinfo {volume}
  {349}},\ \bibinfo {pages} {613} (\bibinfo {year}
  {2015}{\natexlab{a}})}\BibitemShut {NoStop}%
\bibitem [{\citenamefont {Xu}\ \emph {et~al.}(2015{\natexlab{b}})\citenamefont
  {Xu} \emph {et~al.}}]{Xue1501092}%
  \BibitemOpen
  \bibfield  {author} {\bibinfo {author} {\bibfnamefont {S.-Y.}\ \bibnamefont
  {Xu}} \emph {et~al.},\ }\href {\doibase 10.1126/sciadv.1501092} {\bibfield
  {journal} {\bibinfo  {journal} {Science Advances}\ }\textbf {\bibinfo
  {volume} {1}} (\bibinfo {year} {2015}{\natexlab{b}}),\
  10.1126/sciadv.1501092}\BibitemShut {NoStop}%
\bibitem [{\citenamefont {Li}\ \emph {et~al.}(2016{\natexlab{a}})\citenamefont
  {Li} \emph {et~al.}}]{Li:2014bha}%
  \BibitemOpen
  \bibfield  {author} {\bibinfo {author} {\bibfnamefont {Q.}~\bibnamefont {Li}}
  \emph {et~al.},\ }\href {\doibase 10.1038/nphys3648} {\bibfield  {journal}
  {\bibinfo  {journal} {Nature Phys.}\ }\textbf {\bibinfo {volume} {12}},\
  \bibinfo {pages} {550} (\bibinfo {year} {2016}{\natexlab{a}})},\ \Eprint
  {http://arxiv.org/abs/1412.6543} {arXiv:1412.6543 [cond-mat.str-el]}
  \BibitemShut {NoStop}%
\bibitem [{\citenamefont {Manuel}\ and\ \citenamefont
  {Torres-Rincon}(2015)}]{Manuel:2015zpa}%
  \BibitemOpen
  \bibfield  {author} {\bibinfo {author} {\bibfnamefont {C.}~\bibnamefont
  {Manuel}}\ and\ \bibinfo {author} {\bibfnamefont {J.~M.}\ \bibnamefont
  {Torres-Rincon}},\ }\href {\doibase 10.1103/PhysRevD.92.074018} {\bibfield
  {journal} {\bibinfo  {journal} {Phys. Rev.}\ }\textbf {\bibinfo {volume}
  {D92}},\ \bibinfo {pages} {074018} (\bibinfo {year} {2015})},\ \Eprint
  {http://arxiv.org/abs/1501.07608} {arXiv:1501.07608 [hep-ph]} \BibitemShut
  {NoStop}%
\bibitem [{\citenamefont {Ruggieri}\ \emph {et~al.}(2016)\citenamefont
  {Ruggieri}, \citenamefont {Peng},\ and\ \citenamefont
  {Chernodub}}]{Ruggieri:2016asg}%
  \BibitemOpen
  \bibfield  {author} {\bibinfo {author} {\bibfnamefont {M.}~\bibnamefont
  {Ruggieri}}, \bibinfo {author} {\bibfnamefont {G.~X.}\ \bibnamefont {Peng}},
  \ and\ \bibinfo {author} {\bibfnamefont {M.}~\bibnamefont {Chernodub}},\
  }\href {\doibase 10.1103/PhysRevD.94.054011} {\bibfield  {journal} {\bibinfo
  {journal} {Phys. Rev.}\ }\textbf {\bibinfo {volume} {D94}},\ \bibinfo {pages}
  {054011} (\bibinfo {year} {2016})},\ \Eprint
  {http://arxiv.org/abs/1606.03287} {arXiv:1606.03287 [hep-ph]} \BibitemShut
  {NoStop}%
\bibitem [{\citenamefont {Li}\ \emph {et~al.}(2015)\citenamefont {Li} \emph
  {et~al.}}]{Li2015}%
  \BibitemOpen
  \bibfield  {author} {\bibinfo {author} {\bibfnamefont {C.-Z.}\ \bibnamefont
  {Li}} \emph {et~al.},\ }\href {http://dx.doi.org/10.1038/ncomms10137}
  {\bibfield  {journal} {\bibinfo  {journal} {Nat. Commun.}\ }\textbf {\bibinfo
  {volume} {6}},\ \bibinfo {pages} {10137 EP } (\bibinfo {year}
  {2015})}\BibitemShut {NoStop}%
\bibitem [{\citenamefont {Li}\ \emph {et~al.}(2016{\natexlab{b}})\citenamefont
  {Li} \emph {et~al.}}]{Li2016}%
  \BibitemOpen
  \bibfield  {author} {\bibinfo {author} {\bibfnamefont {H.}~\bibnamefont {Li}}
  \emph {et~al.},\ }\href {http://dx.doi.org/10.1038/ncomms10301} {\bibfield
  {journal} {\bibinfo  {journal} {Nat. Commun.}\ }\textbf {\bibinfo {volume}
  {7}},\ \bibinfo {pages} {10301 EP } (\bibinfo {year}
  {2016}{\natexlab{b}})}\BibitemShut {NoStop}%
\bibitem [{\citenamefont {Buividovich}\ \emph {et~al.}(2010)\citenamefont
  {Buividovich}, \citenamefont {Chernodub}, \citenamefont {Kharzeev},
  \citenamefont {Kalaydzhyan}, \citenamefont {Luschevskaya},\ and\
  \citenamefont {Polikarpov}}]{Buividovich:2010tn}%
  \BibitemOpen
  \bibfield  {author} {\bibinfo {author} {\bibfnamefont {P.~V.}\ \bibnamefont
  {Buividovich}}, \bibinfo {author} {\bibfnamefont {M.~N.}\ \bibnamefont
  {Chernodub}}, \bibinfo {author} {\bibfnamefont {D.~E.}\ \bibnamefont
  {Kharzeev}}, \bibinfo {author} {\bibfnamefont {T.}~\bibnamefont
  {Kalaydzhyan}}, \bibinfo {author} {\bibfnamefont {E.~V.}\ \bibnamefont
  {Luschevskaya}}, \ and\ \bibinfo {author} {\bibfnamefont {M.~I.}\
  \bibnamefont {Polikarpov}},\ }\href {\doibase 10.1103/PhysRevLett.105.132001}
  {\bibfield  {journal} {\bibinfo  {journal} {Phys. Rev. Lett.}\ }\textbf
  {\bibinfo {volume} {105}},\ \bibinfo {pages} {132001} (\bibinfo {year}
  {2010})},\ \Eprint {http://arxiv.org/abs/1003.2180} {arXiv:1003.2180
  [hep-lat]} \BibitemShut {NoStop}%
\bibitem [{\citenamefont {Braguta}\ \emph {et~al.}(2016)\citenamefont {Braguta}
  \emph {et~al.}}]{Braguta:2016vhm}%
  \BibitemOpen
  \bibfield  {author} {\bibinfo {author} {\bibfnamefont {V.~V.}\ \bibnamefont
  {Braguta}} \emph {et~al.},\ }\href {\doibase 10.1103/PhysRevB.94.205147}
  {\bibfield  {journal} {\bibinfo  {journal} {Phys. Rev.}\ }\textbf {\bibinfo
  {volume} {B94}},\ \bibinfo {pages} {205147} (\bibinfo {year} {2016})},\
  \Eprint {http://arxiv.org/abs/1608.07162} {arXiv:1608.07162
  [cond-mat.str-el]} \BibitemShut {NoStop}%
\bibitem [{\citenamefont {Braguta}\ \emph {et~al.}(2017)\citenamefont
  {Braguta}, \citenamefont {Katsnelson},\ and\ \citenamefont
  {Kotov}}]{Braguta:2017voo}%
  \BibitemOpen
  \bibfield  {author} {\bibinfo {author} {\bibfnamefont {V.~V.}\ \bibnamefont
  {Braguta}}, \bibinfo {author} {\bibfnamefont {M.~I.}\ \bibnamefont
  {Katsnelson}}, \ and\ \bibinfo {author} {\bibfnamefont {A.~{\relax Yu}.}\
  \bibnamefont {Kotov}},\ }\href@noop {} {\  (\bibinfo {year} {2017})},\
  \Eprint {http://arxiv.org/abs/1704.07132} {arXiv:1704.07132
  [cond-mat.str-el]} \BibitemShut {NoStop}%
\bibitem [{\citenamefont {Montvay}\ and\ \citenamefont
  {Munster}(1997)}]{Montvay:1994cy}%
  \BibitemOpen
  \bibfield  {author} {\bibinfo {author} {\bibfnamefont {I.}~\bibnamefont
  {Montvay}}\ and\ \bibinfo {author} {\bibfnamefont {G.}~\bibnamefont
  {Munster}},\ }\href
  {http://www.cambridge.org/uk/catalogue/catalogue.asp?isbn=0521404320} {\emph
  {\bibinfo {title} {{Quantum fields on a lattice}}}}\ (\bibinfo  {publisher}
  {Cambridge University Press},\ \bibinfo {year} {1997})\BibitemShut {NoStop}%
\bibitem [{\citenamefont {Drut}\ and\ \citenamefont
  {Lahde}(2009{\natexlab{a}})}]{Drut:2008rg}%
  \BibitemOpen
  \bibfield  {author} {\bibinfo {author} {\bibfnamefont {J.~E.}\ \bibnamefont
  {Drut}}\ and\ \bibinfo {author} {\bibfnamefont {T.~A.}\ \bibnamefont
  {Lahde}},\ }\href {\doibase 10.1103/PhysRevLett.102.026802} {\bibfield
  {journal} {\bibinfo  {journal} {Phys. Rev. Lett.}\ }\textbf {\bibinfo
  {volume} {102}},\ \bibinfo {pages} {026802} (\bibinfo {year}
  {2009}{\natexlab{a}})},\ \Eprint {http://arxiv.org/abs/0807.0834}
  {arXiv:0807.0834 [cond-mat.str-el]} \BibitemShut {NoStop}%
\bibitem [{\citenamefont {Drut}\ and\ \citenamefont
  {Lahde}(2009{\natexlab{b}})}]{Drut:2009aj}%
  \BibitemOpen
  \bibfield  {author} {\bibinfo {author} {\bibfnamefont {J.~E.}\ \bibnamefont
  {Drut}}\ and\ \bibinfo {author} {\bibfnamefont {T.~A.}\ \bibnamefont
  {Lahde}},\ }\href {\doibase 10.1103/PhysRevB.79.165425} {\bibfield  {journal}
  {\bibinfo  {journal} {Phys. Rev.}\ }\textbf {\bibinfo {volume} {B79}},\
  \bibinfo {pages} {165425} (\bibinfo {year} {2009}{\natexlab{b}})},\ \Eprint
  {http://arxiv.org/abs/0901.0584} {arXiv:0901.0584 [cond-mat.str-el]}
  \BibitemShut {NoStop}%
\bibitem [{\citenamefont {Hands}\ and\ \citenamefont
  {Strouthos}(2008)}]{Hands:2008id}%
  \BibitemOpen
  \bibfield  {author} {\bibinfo {author} {\bibfnamefont {S.}~\bibnamefont
  {Hands}}\ and\ \bibinfo {author} {\bibfnamefont {C.}~\bibnamefont
  {Strouthos}},\ }\href {\doibase 10.1103/PhysRevB.78.165423} {\bibfield
  {journal} {\bibinfo  {journal} {Phys. Rev.}\ }\textbf {\bibinfo {volume}
  {B78}},\ \bibinfo {pages} {165423} (\bibinfo {year} {2008})},\ \Eprint
  {http://arxiv.org/abs/0806.4877} {arXiv:0806.4877 [cond-mat.str-el]}
  \BibitemShut {NoStop}%
\bibitem [{\citenamefont {Ulybyshev}\ \emph {et~al.}(2013)\citenamefont
  {Ulybyshev} \emph {et~al.}}]{Ulybyshev:2013swa}%
  \BibitemOpen
  \bibfield  {author} {\bibinfo {author} {\bibfnamefont {M.~V.}\ \bibnamefont
  {Ulybyshev}} \emph {et~al.},\ }\href {\doibase
  10.1103/PhysRevLett.111.056801} {\bibfield  {journal} {\bibinfo  {journal}
  {Phys. Rev. Lett.}\ }\textbf {\bibinfo {volume} {111}},\ \bibinfo {pages}
  {056801} (\bibinfo {year} {2013})},\ \Eprint {http://arxiv.org/abs/1304.3660}
  {arXiv:1304.3660 [cond-mat.str-el]} \BibitemShut {NoStop}%
\bibitem [{\citenamefont {DeTar}\ \emph {et~al.}(2017)\citenamefont {DeTar},
  \citenamefont {Winterowd},\ and\ \citenamefont
  {Zafeiropoulos}}]{DeTar:2016dmj}%
  \BibitemOpen
  \bibfield  {author} {\bibinfo {author} {\bibfnamefont {C.}~\bibnamefont
  {DeTar}}, \bibinfo {author} {\bibfnamefont {C.}~\bibnamefont {Winterowd}}, \
  and\ \bibinfo {author} {\bibfnamefont {S.}~\bibnamefont {Zafeiropoulos}},\
  }\href {\doibase 10.1103/PhysRevB.95.165442} {\bibfield  {journal} {\bibinfo
  {journal} {Phys. Rev.}\ }\textbf {\bibinfo {volume} {B95}},\ \bibinfo {pages}
  {165442} (\bibinfo {year} {2017})},\ \Eprint
  {http://arxiv.org/abs/1608.00666} {arXiv:1608.00666 [hep-lat]} \BibitemShut
  {NoStop}%
\bibitem [{\citenamefont {DeTar}\ \emph {et~al.}(2016)\citenamefont {DeTar},
  \citenamefont {Winterowd},\ and\ \citenamefont
  {Zafeiropoulos}}]{DeTar:2016vhr}%
  \BibitemOpen
  \bibfield  {author} {\bibinfo {author} {\bibfnamefont {C.}~\bibnamefont
  {DeTar}}, \bibinfo {author} {\bibfnamefont {C.}~\bibnamefont {Winterowd}}, \
  and\ \bibinfo {author} {\bibfnamefont {S.}~\bibnamefont {Zafeiropoulos}},\
  }\href {\doibase 10.1103/PhysRevLett.117.266802} {\bibfield  {journal}
  {\bibinfo  {journal} {Phys. Rev. Lett.}\ }\textbf {\bibinfo {volume} {117}},\
  \bibinfo {pages} {266802} (\bibinfo {year} {2016})},\ \Eprint
  {http://arxiv.org/abs/1607.03137} {arXiv:1607.03137 [hep-lat]} \BibitemShut
  {NoStop}%
\bibitem [{\citenamefont {Buividovich}\ and\ \citenamefont
  {Polikarpov}(2012)}]{Buividovich:2012nx}%
  \BibitemOpen
  \bibfield  {author} {\bibinfo {author} {\bibfnamefont {P.~V.}\ \bibnamefont
  {Buividovich}}\ and\ \bibinfo {author} {\bibfnamefont {M.~I.}\ \bibnamefont
  {Polikarpov}},\ }\href {\doibase 10.1103/PhysRevB.86.245117} {\bibfield
  {journal} {\bibinfo  {journal} {Phys. Rev.}\ }\textbf {\bibinfo {volume}
  {B86}},\ \bibinfo {pages} {245117} (\bibinfo {year} {2012})},\ \Eprint
  {http://arxiv.org/abs/1206.0619} {arXiv:1206.0619 [cond-mat.str-el]}
  \BibitemShut {NoStop}%
\bibitem [{\citenamefont {Boyda}\ \emph {et~al.}(2016)\citenamefont {Boyda}
  \emph {et~al.}}]{Boyda:2016emg}%
  \BibitemOpen
  \bibfield  {author} {\bibinfo {author} {\bibfnamefont {D.~L.}\ \bibnamefont
  {Boyda}} \emph {et~al.},\ }\href {\doibase 10.1103/PhysRevB.94.085421}
  {\bibfield  {journal} {\bibinfo  {journal} {Phys. Rev.}\ }\textbf {\bibinfo
  {volume} {B94}},\ \bibinfo {pages} {085421} (\bibinfo {year} {2016})},\
  \Eprint {http://arxiv.org/abs/1601.05315} {arXiv:1601.05315
  [cond-mat.str-el]} \BibitemShut {NoStop}%
\bibitem [{\citenamefont {Yamamoto}(2016)}]{Yamamoto:2016rfr}%
  \BibitemOpen
  \bibfield  {author} {\bibinfo {author} {\bibfnamefont {A.}~\bibnamefont
  {Yamamoto}},\ }\href {\doibase 10.1103/PhysRevLett.117.052001} {\bibfield
  {journal} {\bibinfo  {journal} {Phys. Rev. Lett.}\ }\textbf {\bibinfo
  {volume} {117}},\ \bibinfo {pages} {052001} (\bibinfo {year} {2016})},\
  \Eprint {http://arxiv.org/abs/1604.08424} {arXiv:1604.08424 [hep-lat]}
  \BibitemShut {NoStop}%
\bibitem [{\citenamefont {Yamamoto}\ and\ \citenamefont
  {Kimura}(2016)}]{Yamamoto:2016zpx}%
  \BibitemOpen
  \bibfield  {author} {\bibinfo {author} {\bibfnamefont {A.}~\bibnamefont
  {Yamamoto}}\ and\ \bibinfo {author} {\bibfnamefont {T.}~\bibnamefont
  {Kimura}},\ }\href {\doibase 10.1103/PhysRevB.94.245112} {\bibfield
  {journal} {\bibinfo  {journal} {Phys. Rev.}\ }\textbf {\bibinfo {volume}
  {B94}},\ \bibinfo {pages} {245112} (\bibinfo {year} {2016})},\ \Eprint
  {http://arxiv.org/abs/1610.02154} {arXiv:1610.02154 [cond-mat.mes-hall]}
  \BibitemShut {NoStop}%
\bibitem [{\citenamefont {Al-Hashimi}\ and\ \citenamefont
  {Wiese}(2009)}]{AlHashimi:2008hr}%
  \BibitemOpen
  \bibfield  {author} {\bibinfo {author} {\bibfnamefont {M.~H.}\ \bibnamefont
  {Al-Hashimi}}\ and\ \bibinfo {author} {\bibfnamefont {U.~J.}\ \bibnamefont
  {Wiese}},\ }\href {\doibase 10.1016/j.aop.2008.07.006} {\bibfield  {journal}
  {\bibinfo  {journal} {Annals Phys.}\ }\textbf {\bibinfo {volume} {324}},\
  \bibinfo {pages} {343} (\bibinfo {year} {2009})},\ \Eprint
  {http://arxiv.org/abs/0807.0630} {arXiv:0807.0630 [quant-ph]} \BibitemShut
  {NoStop}%
\bibitem [{\citenamefont {DeGrand}\ and\ \citenamefont
  {DeTar}(2006)}]{DeGrand:2006zz}%
  \BibitemOpen
  \bibfield  {author} {\bibinfo {author} {\bibfnamefont {T.}~\bibnamefont
  {DeGrand}}\ and\ \bibinfo {author} {\bibfnamefont {C.~E.}\ \bibnamefont
  {DeTar}},\ }\href@noop {} {\emph {\bibinfo {title} {{Lattice methods for
  quantum chromodynamics}}}}\ (\bibinfo {year} {2006})\BibitemShut {NoStop}%
\bibitem [{\citenamefont {Kharzeev}\ \emph
  {et~al.}(2013{\natexlab{b}})\citenamefont {Kharzeev}, \citenamefont
  {Landsteiner}, \citenamefont {Schmitt},\ and\ \citenamefont
  {Yee}}]{Kharzeev:2013jha}%
  \BibitemOpen
  \bibfield  {author} {\bibinfo {author} {\bibfnamefont {D.}~\bibnamefont
  {Kharzeev}}, \bibinfo {author} {\bibfnamefont {K.}~\bibnamefont
  {Landsteiner}}, \bibinfo {author} {\bibfnamefont {A.}~\bibnamefont
  {Schmitt}}, \ and\ \bibinfo {author} {\bibfnamefont {H.-U.}\ \bibnamefont
  {Yee}},\ }\href {\doibase 10.1007/978-3-642-37305-3} {\bibfield  {journal}
  {\bibinfo  {journal} {Lect. Notes Phys.}\ }\textbf {\bibinfo {volume}
  {871}},\ \bibinfo {pages} {pp.1} (\bibinfo {year}
  {2013}{\natexlab{b}})}\BibitemShut {NoStop}%
\end{thebibliography}

\end{document}